\documentclass[10pt,aps,pra,twocolumn,superscriptaddress,floatfix,nofootinbib]{revtex4-2}

\usepackage{amsmath, amsfonts,mathrsfs, amssymb,mathtools}
\usepackage{bm,bbm, braket, accents}
\usepackage{graphicx}   
\usepackage[usenames,dvipsnames]{color}
\usepackage{algpseudocode,algorithm,algorithmicx}

\makeatletter
\def\@bibdataout@aps{%
 \immediate\write\@bibdataout{%
  @CONTROL{%
   apsrev41Control,author="08",editor="1",pages="0",title="0",year="1",eprint="1"%
  }%
 }%
 \if@filesw
  \immediate\write\@auxout{\string\citation{apsrev41Control}}%
 \fi
}%
\makeatother 

\newcommand{\sch}{Schr\"odinger }

\newcommand{\iprod}[2]{\left\langle{#1}\middle\vert{#2}\right\rangle}
\newcommand{\oprod}[2]{\left\vert{#1}\middle\rangle\!\middle\langle{#2}\right\vert}

\newcommand{\proj}[1]{\hat P_{{#1}}}
\newcommand{\calproj}[1]{P_{{#1}}}

\usepackage[breaklinks=true]{hyperref}
\hypersetup{
  colorlinks   = true, 
  urlcolor     = blue, 
  linkcolor    = blue, 
  citecolor   = red 
}

\usepackage[dvipsnames]{xcolor}

\usepackage[deletedmarkup=sout, authormarkup=superscript]{changes}
\definechangesauthor[color=Red]{ToDo}
\definechangesauthor[color=Plum]{MN}
\definechangesauthor[color=ForestGreen]{JC}

\newcommand\change{}

\usepackage[capitalise]{cleveref} 
\crefformat{equation}{Eq.~(#2#1#3)} 
\crefformat{section}{Sec.~#2#1#3} 
\Crefformat{equation}{Equation~(#2#1#3)}
\crefformat{figure}{Fig.~#2#1#3}
\crefrangeformat{equation}{Eqs.~#3(#1)#4--#5(#2)#6}
\Crefformat{section}{Section~#2#1#3}

\begin{document}

\title{A Projection Operator-based Newton Method for the\\ Trajectory Optimization of Closed Quantum Systems}

\author{Jieqiu Shao}
\affiliation{Department of Electrical, Computer and Energy Engineering, University of Colorado Boulder, Boulder, Colorado 80309, USA}

\author{Joshua Combes}
\affiliation{Department of Electrical, Computer and Energy Engineering, University of Colorado Boulder, Boulder, Colorado 80309, USA}

\author{John Hauser}
\affiliation{Department of Electrical, Computer and Energy Engineering, University of Colorado Boulder, Boulder, Colorado 80309, USA}

\author{Marco M. Nicotra}
\affiliation{Department of Electrical, Computer and Energy Engineering, University of Colorado Boulder, Boulder, Colorado 80309, USA}

\date{\today}

\begin{abstract}
Quantum optimal control is an important technology that enables fast state preparation and gate design. 
In the absence of an analytic solution, most quantum optimal control methods rely on an iterative scheme to update the solution estimate. At present, the convergence rate of existing solvers is, at most, superlinear. 
This paper develops a new general purpose solver for quantum optimal control based on the PRojection Operator Newton method for Trajectory Optimization, or PRONTO. Specifically, the proposed approach uses a projection operator to incorporate the Schr\"odinger equation directly into the cost function, which is then minimized using a Newton descent method. At each iteration, the descent direction is obtained by computing the analytic solution to a linear-quadratic trajectory optimization problem. The resulting method guarantees monotonic convergence at every iteration and quadratic convergence in proximity of the solution. \change{The potential of PRONTO is showcased by solving the optimal state-to-state mapping problem for a qubit and providing comparisons} to a state-of-the-art quantum optimal control method.
\end{abstract}

\maketitle

\section{Introduction}\label{sec:intro}
To accomplish the promise of quantum computing and quantum sensing, it is necessary to reliably and accurately control increasingly large quantum systems.
Quantum control~\cite{Brif2010,Dong2010,Altafini2012,Q_Opt2015} is difficult for many reasons, including: interesting quantum systems are fundamentally nonlinear~\cite{Lloyd_Braun_99}, 
observations disturb the state of the system being controlled, and, like classical systems, the quantum state space suffers from the curse of dimensionality.

Broadly speaking, the field of quantum optimal control can be divided into model-free, e.g. \cite{CRAB2011}, and model-based methods. In the context of model-based quantum optimal control, the two predominant strategies in modern literature are the GRadient Ascent Pulse Engineering algorithm (GRAPE) \cite{GRAPE,Khaneja_Brockett_2001}, which treats the control input as a sequence of piecewise-constant pulses, and Krotov methods \cite{Krotov1,Krotov2}, which treat the control input as a continuous function. For a detailed comparison between the two strategies, readers are referred to \cite{Wilhelm2020}. Because these prior methods use gradient-based descent, their convergence to the optimal solution is predominantly linear (or superlinear in the case of quasi-Newton gradient acceleration schemes \cite{Krotov_qNewton,GRAPE_qNewton}). The development of quadratically convergent quantum control methods is an open problem and may lead to the ability to control larger quantum systems.

This paper introduces a new strategy for model-based quantum optimal control by specializing the PRojection Operator-based Newton method for Trajectory Optimization (PRONTO) \cite{PRONTO2002} to quantum systems. An example featuring a prior application of PRONTO to quantum control is discussed briefly in \cite{qPRONTO}.

PRONTO is conceptually similar to Krotov, in the sense that they both solve the optimal control problem directly in function space, and they both do so by sequentially solving backward-in-time and forward-in-time ordinary differential equations. Their difference lies in how they handle the system dynamics: Krotov enforces them using Lagrange multipliers and employs a primal-dual gradient method to seek the saddle point of the Lagrangian; PRONTO embeds them into a modified cost function using the projection operator, and employs a Newton descent method to seek the minimum of the modified cost function. Thus, the convergence rate of Krotov is linear, whereas the convergence rate of PRONTO can be quadratic.

The paper is structured as follows: Section \ref{sec:prob_statement} states the quantum optimal control problem we wish to solve and reformulates it into a traditional optimal control problem. Section \ref{sec:Prelim} reviews established results from numerical optimization theory to familiarize the reader with the general concepts used in PRONTO. Section \ref{sec:PRONTO} provides a detailed description of how to implement PRONTO on quantum systems and includes a pseudo-code summarizing all the steps performed by the method. Finally, Section \ref{sec:NumEx} provides preliminary comparisons between PRONTO and the state-of-the-art Krotov implementation package \cite{KrotovToolkit2019}.

\section{Problem Statement}\label{sec:prob_statement}
The objective of this paper is to develop a systematic approach for the state-to-state control of quantum systems, meaning that we wish to steer some initial state $\ket{\psi_0}$ to a target state $\ket{\phi}$ over a finite time interval $[0,T]$. In this section, we show how the state-to-state quantum control problem can be reformulated as a classic optimal control problem, which we will then solve using a specialized version of the PRONTO algorithm.

\subsection{Dynamic Model}
We consider a system governed by the Hamiltonian
\begin{equation}
    \hat H[u(t)] = \hat H_0 + \sum_{j=1}^m\hat H_j\, f_j[u_j(t)],
\end{equation}
where $\hat H_0\in\mathbb C^{n\times n}$ represents the free evolution of the system, $\hat H_j\in\mathbb C^{n\times n}$ is the $j$-th control Hamiltonian and the associated $f_j:\mathbb R \to \mathbb R$ is a class $\mathcal{C}^2$ function of the control input $u_j(t)\in \mathbb R$. The dynamics generated by this Hamiltonian are 
\begin{equation}\label{eq:sch_eq1}
     i\hbar\:\frac{\partial}{\partial t}\ket{\psi(t)} =  \hat H[u(t)] \ket{\psi(t)},
\end{equation}
with $\ket{\psi(t)}\in\mathbb C^n$ and $u(t)\in \mathbb R^m$. 
Henceforth, we set $\hbar=1$ for simplicity. 

\subsection{Cost Function}

To quantify the effectiveness of a given controller, we will rank the behavior of $(\ket{\psi(t)}$, $u(t))$  within the time window $t\in[0,T]$, by evaluating the cost function
\begin{equation}\label{eq:ocp_Qcost}
    \hat m[\,\ket{\psi(T)}]+\int_0^T \hat l [\,\ket{\psi(t)},u(t)]dt,
\end{equation}
where both the {\em terminal cost} $\hat m:\mathbb C^n\to\mathbb R_{\geq0}$ and the {\em incremental cost} $\hat l:(\mathbb C^n\times\mathbb R^m)\to\mathbb R_{\geq0}$ are class $\mathcal C^2$ convex functions that can be interpreted as follows: \medskip 

$\bullet~$The incremental cost penalizes undesirable behaviors within the time window $t\in[0,T]$. A common choice is 
\begin{equation}\label{eq:qincrimental_cost}
    \hat l (\ket{\psi},u\big)= \hat l_\psi(\ket\psi) + l_u(u),
\end{equation}
with $\hat l_\psi$ convex and $l_u$ strongly convex. The latter prevents $u(t)$ from growing unbounded, and is therefore necessary for the well-posedness of the problem. A typical choice for $l_u\neq0$ is
\begin{equation}
    l_u(u)=\tfrac12u^T R(t)\, u.
\end{equation}
where $R(t)>0,~\forall t\in[0,T]$ is a (potentially time-varying) symmetric matrix. As for the former, $\hat l_\psi\neq0$ can be useful for penalizing the transfer of population onto undesirable states $\ket\lambda$, as detailed in \cite{Palao2008}. In this case, a suitable choice would be
\begin{equation}
    \hat l_\psi(\ket\psi)=\tfrac12\bra{\psi}\proj{\lambda}\ket{\psi},
\end{equation}
where the operator $\proj{\lambda} = \oprod{\lambda}{\lambda}$ is a projection of $\ket\psi$ onto a state $\ket{\lambda}$. If we do not wish to penalize any states during the transient, it is worth noting that $\hat l_\psi=0$ is a suitable, and fairly common \cite{KrotovToolkit2019}, alternative. \medskip

$\bullet~$ The terminal cost drives the final state $\ket{\psi(T)}$ to a desirable target $\ket\phi$ by penalizing the deviation between $\ket{\psi(T)}$ and $\ket\phi$. For example, we can perform an arbitrary phase state-to-state transition by assigning 
\begin{equation}\label{eq:qterminal_cost}
\hat m(\ket\psi)= \bra{\psi}\proj{\neg\phi}\ket{\psi},
\end{equation}
where the operator $\proj{\neg \phi} = 1-\oprod\phi\phi$ is the complement of $\proj{\phi}$ and $\hat m$ denotes the squared Hilbert-Schmidt distance between $\ket\psi$ and the target state $\ket{\phi}$.

\subsection{Optimal Control Problem}

Having identified the system dynamics and a suitable cost function, we can now formulate the well-known \cite{Brif2010} quantum optimal control problem
\begin{subequations}\label{eq:ocp_Quantum}
\begin{eqnarray}
\displaystyle \min ~&&\displaystyle \hat m[\,\ket{\psi(T)}]+\int_0^T \hat l[\,\ket{\psi(t)},u(t)]dt\\
\textrm{s.t.}~&& 
\tfrac\partial{\partial t}\ket{\psi(t)} \!=\! -i\hat H[u(t)]\!\ket{\psi(t)}\!,~ \ket{\psi(0)}\!=\!\ket{\psi_0}\!.
\end{eqnarray}
\end{subequations}
To solve this problem, we will start by re-framing it into the form found in conventional control literature \cite{Opt_ctrl}. This is done by
transforming the complex state vector $\ket{\psi(t)}\in \mathbb{C}^n$ into a larger real vector $x(t) \in \mathbb{R}^{2n}$ by taking advantage of the bijective mapping 
\begin{equation}\label{eq:bijective}
    x(t)=\begin{bmatrix}
    \textrm{Re}\big [\ket{\psi(t)}\big ]\\\textrm{Im}    \big [\ket{\psi(t)}\big ]
    \end{bmatrix}.
\end{equation}
This mapping, detailed in \cite{BijectiveMapping}, allows us to transform any operator $\hat Y\in\mathbb C^{n\times n}$ into a matrix $Y\in\mathbb R^{2n\times2n}$ using 
\begin{equation}\label{eq:operator_bijective}
    Y=\begin{bmatrix}\textrm{Re}(\hat Y)&-\textrm{Im}(\hat Y)~\\ \textrm{Im}(\hat Y) & ~~\textrm{Re}(\hat Y)
    \end{bmatrix}.
    \end{equation}
The Schr\"odinger equation \eqref{eq:sch_eq1} can therefore be rewritten as the real-valued ordinary differential equation 
\begin{equation}
    \dot x(t)= H[u(t)]\,x(t)
\end{equation}
by defining
\begin{equation}\label{eq:H_mapping}
     H =\begin{bmatrix}\textrm{Re}(-i \hat H)&-\textrm{Im}(-i\hat H)~\\ \textrm{Im}(-i\hat H) & ~~\textrm{Re}(-i\hat H)~
    \end{bmatrix}.
\end{equation}
Likewise, \change{the bijective mapping \eqref{eq:bijective} can be leveraged to transform the functions $\hat l(\ket \psi,u)$ and $\hat m(\ket\psi)$ into their equivalent form $l(x,u)$ and $m(x)$. For example,} the cost functions \eqref{eq:qincrimental_cost}-\eqref{eq:qterminal_cost} can be rewritten as
\begin{subequations}
\begin{eqnarray}
    l(x,u)&&= \tfrac12 x^T \calproj{\lambda}\, x+ 
    \tfrac12 u^T R(t)\, u,\\
    m(x)&&= \tfrac12 x^T\calproj{\neg \phi}\, x,
\end{eqnarray}
\end{subequations}
where the matrices $\calproj{\neg\phi}$ and $\calproj{\lambda}$ can be obtained from the operators $\proj{\neg\phi}$ and $\proj{\lambda}$ using \cref{eq:operator_bijective}.\medskip

Solving the quantum optimal control problem \eqref{eq:ocp_Quantum} is therefore equivalent to solving the conventional optimal control problem 
\begin{subequations}\label{eq:ocp_original}
\begin{eqnarray}
\displaystyle \min \quad&&\displaystyle m[x(T)]+\int_0^Tl[x(t),u(t)]dt \label{eq:opt_cost}\\
\textrm{s.t.}\quad&& \dot x(t)=H[u(t)]\,x(t),\quad x(0)= x_0\label{eq:opt_cstr}.
\end{eqnarray}
\end{subequations}
Compared to the more general formulation featuring $\dot x = f(x,u)$, \eqref{eq:ocp_original} features two very interesting properties that stem from the nature of closed quantum systems
\begin{itemize}
    \item The differential equation \eqref{eq:opt_cstr} is \emph{affine} in the state vector $x$, meaning that its second derivative satisfies $\nabla^2_{xx}[H(u)\,x]=0$;
    \item The matrix $H(u)$ is \emph{skew-symmetric} for all $u\in\mathbb R^m$. This implies $\|x(t)\|=\|x(0)\|,~\forall t\in[0,T]$.
\end{itemize}
These properties, plus the general simplicity of computing the partial derivatives of \eqref{eq:ocp_original} will be leveraged in \cref{sec:PRONTO} to design an efficient iterative solver.




\section{Optimization primer}\label{sec:Prelim}
The goal of this section is to provide the interested reader with an intuition for the main ingredients used in PRONTO.
Although most of the information contained in this section can be found in a good textbook on numerical optimization, e.g. \cite{Nocedal}, the order in which it is presented and the way it is interpreted is both original to this paper and critical for understanding the theory behind PRONTO.

The starting point is \cref{eq:ocp_original}, which is a constrained optimization problem defined in \emph{function space}, meaning that its solution $[x(t),u(t)]$ is a pair of functions $x:[0,T]\to\mathbb R^{2n}$ and $u:[0,T]\to\mathbb R^m$. Since readers are more likely familiar with \emph{vector space} optimization, this section describes the general intuition behind the approach used in \cref{sec:PRONTO} for the simplified case where $x\in\mathbb R^{2n}$ and $u\in\mathbb R^m$ are just two vectors.

\subsection{Embedding Constraints into the Cost Function}

Given a constrained optimization problem in the form
\begin{subequations}
\begin{eqnarray}
\displaystyle \min~~ && h(x,u) \\
\textrm{s.t.}~~&& c(x,u)=0,
\end{eqnarray}
\end{subequations}
where $h:(\mathbb R^{2n}\times\mathbb R^m)\to\mathbb R_{\geq0}$ is a $\mathcal{C}^2$ convex function and $c:(\mathbb R^{2n}\times\mathbb R^m)\to\mathbb R$ is a $\mathcal{C}^2$ function, a systematic method for finding the solution is to search for the saddle point of the Lagrangian $\mathcal{L}(x,u,\chi)=h(x,u)+\chi^T[c(x,u)]$, where $\chi\in\mathbb R^{2n}$ is the vector of Lagrange multipliers. This can be done using a primal-dual gradient method (incidentally, if one were to transfer this intuition back into function space, one would obtain something akin to the Krotov method \cite{Krotov_orig}).

Given a $\mathcal{C}^2$ function $p:\mathbb R^m\to(\mathbb R^{2n},\mathbb R^m)$ such that $c(x,u)=0$ iff $[x,u]=p(u)$, an alternative method would be to solve the unconstrained optimization problem 
\begin{equation}\label{eq:ex_ocp}
    \min~g(u),
\end{equation}
with $g(u)=h[p(u)]$. Clearly, the main challenge associated to this transformation is finding a suitable function $p(u)$. Doing so, however, enables us to use a standard Newton method for finding the solution.


\subsection{Newton Method}\label{ssec:Newton_ex}
The classic Newton method seeks local minima of \eqref{eq:ex_ocp} by applying an iterative procedure in the form
\begin{equation}\label{eq:ex_cN}
    u_{k+1}=u_k+\nu_k,
\end{equation}
where, given a guess $u_k$, the \textbf{Descent Direction} $\bm{\nu_k}$ is obtained by computing the local quadratic approximation of the cost update, i.e.
\begin{equation}\label{eq:ex_2oA}
    g(u)\approx g(u_k)+\nabla^T g(u_k)\nu+\tfrac12\nu^T\nabla^2 g(u_k)\nu,
\end{equation}
and solving a quadratic minimization problem
\begin{equation}\label{eq:ex_LQR}
    \nu_k=\arg\min \nabla^Tg(u_k)\nu+\tfrac12\nu^T\nabla^2 g(u_k)\nu.
\end{equation}
The interest in this approach is that, unlike the original problem \eqref{eq:ex_ocp}, the local quadratic approximation \eqref{eq:ex_LQR} can be solved explicitly iff $\nabla^2 g(u_k)>0$. Notably, the solution to \eqref{eq:ex_LQR} is
\begin{equation}
    \nu_k=-\nabla^2 g(u_k)^{-1}\nabla g(u_k).
\end{equation}
When a local minimizer of a $\mathcal C^2$ function satisfies  $\nabla^2 g(u^\star)>0$, it can be shown that, given a sufficiently small initial error $\|u_0-u^\star\|$, the sequence $\|u_k-u^\star\|$ is quadratically convergent to zero. Moreover, since a local minimizer of a $\mathcal C^2$ function satisfy  $\nabla g(u^\star)=0$, it is possible to use $\|\nabla g(u_k)\|\leq \texttt{tol}$, with $\texttt{tol}>0$, as an exit condition for the iterative solver.

\subsection{quasi-Newton Method}
Even in cases where the requirement $\nabla^2 g(u_k)>0$ holds for $u_k$ sufficiently close to the optimum, there is no guarantee that $\nabla^2 g(u_k)>0,~\forall u_k\in\mathbb R^m$. Given $\nabla^2 g(u_k)\not>0$, the quadratic approximation \eqref{eq:ex_LQR} does not admit a bounded solution, which causes the Newton method to fail. In such instances, the quasi-Newton method can overcome this issue by replacing \eqref{eq:ex_LQR} with
\begin{equation}\label{eq:ex_quasiN}
    \nu_k=\arg\min_{\nu\in\mathbb R^m} \nabla g(u_k)\nu+\tfrac12\nu^T G_k \nu,
\end{equation}
where $G_k>0$ is a positive-definite matrix. Since $G_k=I$ leads to the gradient descent method $\nu_k=-\nabla g(u_k)$, which is linearly convergent, we note that, if $G_k>0$ is a suitable approximation of $\nabla^2 g(u_k)$, the quasi-Newton method can achieve superlinear convergence. 

For example, given $g(u)=h[p(u)]$, we have
\[
  \nabla^2 g(u) = \nabla p(u)^T\nabla^2 h[p(u)]\nabla p(u)+\sum_{j=1}^m \left.\frac{\partial h}{\partial u_j}\right|_{p(u)}\!\!\!\!\!\!\!\!\!\!\nabla^2p_j(u)
\]
which is, in general, not positive definite. If $h$ is a convex function, however, a suitable positive definite approximation of $\nabla^2 g(u_k)\not>0$ would be
\begin{equation}
    G_k=\nabla p(u_k)^T\nabla^2 h[p(u_k)]\nabla p(u_k).
\end{equation}


\subsection{Dampened (quasi-)Newton Method}\label{ssec:Armijo}
\begin{figure}
    \centering
    \includegraphics[width=0.48\textwidth]{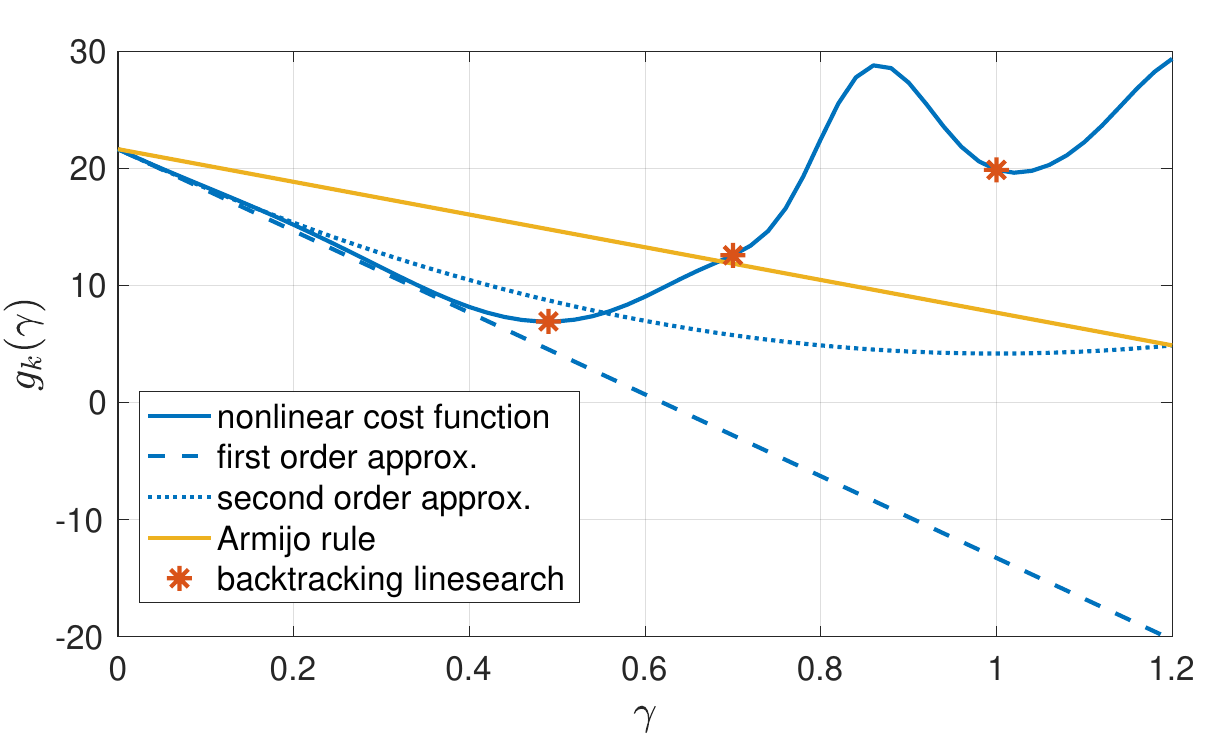}
    \caption{Visual representation of the backtracking linesearch applied to a nonlinear cost function $g_k(\gamma)$. For $\gamma=1$ and $\gamma=0.7$, the nonlinearities are too strong to be captured by a second order approximation. Given $\gamma=0.49$, the Armijo rule is satisfied, meaning that the decrease in cost is acceptable. Note that, for $\gamma\to0$, the nonlinear function $g_k(\gamma)$ tends to coincide with its first and second order approximations.  \label{fig:Armijo}}
\end{figure}

Since \eqref{eq:ex_2oA}, or its quasi-Newton counterpart \eqref{eq:ex_quasiN}, is only a second-order approximation, there is no guarantee that $g(u_k+\nu_k)<g(u_k)$. This can cause the Newton method to diverge when the initial error $\|u_0-u^\star\|$ is too large. To prevent this, we replace the classic Newton step \eqref{eq:ex_cN} with the dampened Newton step
\begin{equation}
    u_{k+1}=u_k+\gamma_k\nu_k,
\end{equation}
where the \textbf{Step Size} $\bm{\gamma_k\in(0,1]}$ is chosen to ensure a sufficient decrease in the cost function. Notably, if we define $g_k:(0,1]\to\mathbb{R}$ as
\begin{equation}
    g_k(\gamma)=g(u_k+\gamma\nu_k),
\end{equation}
we can compute the Taylor expansion
\begin{equation}\label{eq:ex_Taylor}
    g_k(\gamma)=g_k(0)+g_k'(0)\gamma+\frac12g_k''(0)\gamma^2+o(\gamma^2).
\end{equation}
By construction of $\nu_k$, we know that the minimum of the quadratic approximation of $g_k(\gamma)$ lies in $\gamma=1$. As shown in Figure \ref{fig:Armijo}, however, the higher order terms $o(\gamma^2)$ may cause $g_k(1)$ to be quite different from its quadratic approximation, thus causing the Newton method to fail. To ensure that the cost decreases by a sufficient amount at each iteration, we use a backtracking linesearch to enforce the Armijo Rule
\begin{equation}\label{eq:ex_Armijo}
    g_k(\gamma_k)\leq g_k(0)-\alpha g'_k(0)\gamma_k,
\end{equation}
with $\alpha\in(0,0.5)$. Specifically, given the initial value $\gamma_k=1$, we check if \eqref{eq:ex_Armijo} holds. If it does, the higher-order terms $o(\gamma^2)$ are not dominant. If it doesn't, we assign 
\begin{equation}\label{eq:ex_backtrack}
    \gamma_k\leftarrow\beta\gamma_k,
\end{equation}
with $\beta\in(0,1)$ and proceed to once again check the Armijo Rule \eqref{eq:ex_Armijo}. Typical values for the backtracking linesearch are $\alpha=0.4$ and $\beta=0.7$. This modification is sufficient to ensure the monotonic convergence property
\begin{equation}
    g(u_{k+1})<g(u_k),
\end{equation}
thus guaranteeing monotonic convergence to a local minimum for any initial condition such that $\nabla g(u_0)\neq0$, i.e. for any initial condition that isn't a local maximum or a saddle point. 

\section{PRONTO for Quantum Systems}\label{sec:PRONTO}
Having given a general intuition behind of all the steps used in PRONTO, we now go back to the original optimization problem \eqref{eq:ocp_original} for which the solution space is $[x(t),u(t)]\in(\mathcal{X}\times\mathcal{U})$, where $\mathcal{X}$ denotes the set of all $\mathcal{C}^0$ functions $x:[0,T]\to\mathbb R^{2n}$ and  $\mathcal{U}$ denotes the set of all $\mathcal{C}^0$ functions $u:[0,T]\to\mathbb R^{m}$. Here, the idea is to perform the same steps laid out in \cref{sec:Prelim} using functional analysis instead of vector calculus. The PRONTO pseudocode obtained by following all these steps is provided in Algorithm \ref{alg:Q-PRONTO} at the end of this section.

\subsection{The Projection Operator}

The first step to solve the constrained optimization problem \eqref{eq:ocp_original} is to reformulate it as an \emph{unconstrained} optimization problem by embedding the equality constraints into the cost function. To this end, we define $\xi(t)=[x(t),u(t)]$ which means we can rewrite \eqref{eq:opt_cost} as
\begin{equation}\label{eq:cost}
    h(\xi)=m[x(T)]+\int_0^Tl[x(t),u(t)]dt,
\end{equation}
where $h: \mathcal X \times \mathcal U\to\mathbb{R}$ is a $\mathcal{C}^2$ convex function. Since the \sch equation \eqref{eq:opt_cstr} imposes the restriction $\xi\in\mathcal T$, where
\begin{equation}
    \mathcal{T}=\{\xi\in(\mathcal X \times \mathcal U)~|~\dot x=\mathcal H(u)\,x,~~x(0)=x_0\},\vspace{3pt}
\end{equation} 
we define the simplified\footnote{As detailed in \cite{PRONTO2002}, the projection operator should formally map $(\mathcal X\times \mathcal U)\to\mathcal T$. Given $\alpha\in\mathcal X$, we can turn \eqref{eq:simp_Proj} into a ``proper'' projection operator $\mathcal P(\alpha,\mu)$ by defining $u=\mu+K_r(\alpha-x)$, where $K_r$ is a time-varying feedback gain that stabilizes the system trajectories. Since closed quantum systems evolve on the unit sphere (and are therefore inherently stable), we assign $K_r=0$ for the sake of simplicity. The design of suitable $K_r\neq0$, which is likely to improve convergence, is left to future work.} projection operator $\mathcal{P}:\mathcal{U}\to\mathcal T$ as the solution to
\begin{equation}\label{eq:simp_Proj}
    \mathcal{P}(\mu)=\xi:\left\{\begin{array}{l}
     \dot{x}=H(u)\,x, ~~x_0=\bar x\\
         u = \mu.
    \end{array}\right. 
\end{equation}
This operator simply maps a specific control field $\mu(t)$ to an abstract state $\xi(t)$ consisting of both the control field and the dynamic response of the state under that control field. Then, given $g(u)=h(\mathcal{P}(u))$, the optimal control problem \eqref{eq:ocp_original} can be rewritten as an \emph{unconstrained} optimization problem in the form
\begin{equation}\label{eq:simp_ocp}
\min~ g(u),
\end{equation}
which we now solve using a dampened (quasi-)Newton method in the function space $\mathcal{U}$.

\subsection{Newton Descent Direction}
In analogy to \cref{ssec:Newton_ex}, given the current solution estimate $u_k(t)\in\mathcal U$, we now wish to compute a descent direction $\nu_k(t)\in\mathcal{U}$ for our next estimate by minimizing the local quadratic approximation of \eqref{eq:simp_ocp}. This can be done by solving the optimization problem

\begin{equation}\label{eq:simp_LQR}
    \nu_k=\arg\min ~ D g(u_k)\circ \nu +\frac12D^2 g(u_k)\circ(\nu,\nu),
\end{equation}
where $D$ and $D^2$ are the first and second Fr\'echet derivatives of $g(u)$. As detailed in Appendix \ref{app:derivatives}, the evaluation of \eqref{eq:simp_LQR} leads in the following Linear-Quadratic Optimal Control Problem (LQ-OCP) 
\begin{widetext}
\begin{equation}\label{eq:LQR}
\begin{array}{rl}
    \displaystyle \nu_k(t)=\arg\min ~&\displaystyle \pi_k^Tz(T)+\frac12z(T)^T\Pi_kz(T)+ \displaystyle\int_0^T\begin{bmatrix}q_k(\tau)\\r_k(\tau)\end{bmatrix}^T\begin{bmatrix}z(\tau)\\\nu(\tau)\end{bmatrix}+\frac12\begin{bmatrix}z(\tau)\\\nu(\tau)\end{bmatrix}^T\begin{bmatrix}Q_k(\tau) & S_k(\tau)\\S_k^T(\tau)&R_k(\tau)\end{bmatrix}\begin{bmatrix}z(\tau)\\\nu(\tau)\end{bmatrix}
    d\tau\\~\\
\textrm{s.t.}~& \dot z(t)=A_k(t)z(t)+B_k(t)\nu(t),\qquad z(0)=0,
\end{array}
\end{equation}
\end{widetext}
where
\begin{equation}\label{eq:FirstOrder}
\begin{array}{l}
    q_k(t)=\nabla_xl(x_k(t),u_k(t)),\\
    r_k(t)=\nabla_ul(x_k(t),u_k(t)),\\
    \pi_k=\nabla_xm(x_k(T)),
\end{array}
\end{equation}
capture the first-order contributions to the cost and \begin{equation}\label{eq:Newton}
\begin{array}{l}
    Q_k(t)=\nabla^2_{xx}l(x_k(t),u_k(t)),\\
    R_k(t)=\nabla^2_{uu}l(x_k(t),u_k(t))+\tilde R_k(t),\\S_k(t)=\nabla^2_{xu}l(x_k(t),u_k(t))+\tilde S_k(t),\\
    \Pi_k=\nabla^2_{xx}m(x_k(T)),
\end{array}
\end{equation}
capture the second-order contributions, with 
\[
\begin{array}{rl}
    \tilde S_k(t)&=\displaystyle\begin{bmatrix} H_1(t)\chi_k(t)&\ldots& H_m(t)\chi_k(t) \end{bmatrix}\\~\\
    \tilde R_k(t)&=\displaystyle\begin{bmatrix}\chi_k^T(t) H_{11}(t)x_k(t)&\ldots&\chi_k^T(t) H_{m1}(t)x_k(t)\\
    \vdots&\ddots&\vdots\\ \chi_k^T(t) H_{1m}(t)x_k(t)&\ldots&\chi_k^T(t) H_{mm}(t)x_k(t) \end{bmatrix}.
\end{array}
\]
Here, the brackets indicate a matrix concatenation of column vectors for $\tilde S_k$ and scalars for $\tilde R_k$, the co-state\footnote{Coincidentally, it is worth noting that the co-state dynamics \eqref{eq:adjoint}, which stem from the computation of $D^2g(u_k)(\nu,\nu)$, detailed in \cref{app:derivatives}, are identical to what is found by the Krotov method \cite[Eq. (30)]{KrotovToolkit2019} using Lagrange multipliers.} $\chi_k\in\mathcal X$ is obtained by solving the differential equation
\begin{equation}\label{eq:adjoint}
    -\dot\chi_k= H(u_k(t))^T\chi_k+q_k(t),\qquad\chi(T)=\pi_k,
\end{equation}
and, given the vector $w\in\mathbb{R}^m$, we define
\[
\begin{array}{rl}
    H_i(t) = \displaystyle\left.\frac{\partial H(w)}{\partial w_i}\right|_{w=u_k(t)}
    &
    \quad H_{ij}(t) = \displaystyle\left.\frac{\partial^2 H(w)}{\partial w_i\partial w_j}\right|_{w=u_k(t)}.
\end{array}
\]
Finally, the matrices
\begin{equation}\label{eq:LinDyn}
\begin{array}{l}
    A_k(t)=H(u_k(t)),\\
    B_k(t)=\displaystyle\begin{bmatrix} H_1(t)x_k(t)&\ldots& H_m(t)x_k(t) \end{bmatrix},
\end{array}
\end{equation}
capture the linearization of system dynamics $\dot x= H(u)x$ around the current trajectory $\xi_k=(x_k(t),u_k(t))$.\medskip

Although somewhat daunting, \eqref{eq:LQR} is a well-known optimal control problem \cite{Opt_ctrl}, which admits an explicit solution (see \cref{app:LQR}). Unfortunately, depending on the matrices in \eqref{eq:Newton}, there is no guarantee that \eqref{eq:LQR} the solution exists and is unique. This issue is solved in the next subsection by using a positive definite approximation of the cost function whenever necessary.

\subsection{quasi-Newton Descent Direction}
If, at a given iteration $k$, the LQ-OCP \eqref{eq:LQR} does not admit a solution, it is possible to replace \eqref{eq:Newton} with
\begin{equation}\label{eq:quasi_Newton}
\begin{array}{ll}
    Q_k(t)=\nabla^2_{xx}l(x_k(t),u_k(t)),\\
    R_k(t)=\nabla^2_{uu}l(x_k(t),u_k(t)),\\S_k(t)=\nabla^2_{xu}l(x_k(t),u_k(t)),\\
    \Pi_k=\nabla^2_{xx}m(x_k(T)).
\end{array}
\end{equation}
By construction, these matrices satisfy the positive semi-definite properties $\Pi_k\geq0$, $R_k(t)>0$, $\forall t\in[0,T]$, and $[Q_k(t)~S_k(t);S_k(t)^T~R_k(t)]\geq0,~\forall t\in[0,T]$. This is sufficient to ensure that the new LQ-OCP has a unique minimizer, making it always possible to compute a descent direction $\nu_k$. \smallskip

It is very interesting, and somewhat counter-intuitive, to note that the co-state $\chi_k(t)$ is not required to compute the quasi-Newton descent direction since $\chi_k(t)$ only enters the problem through the matrices $\tilde S_k$, $\tilde R_k$ in \eqref{eq:Newton}.

\subsection{Maximum Step Size}

In analogy to \cref{ssec:Armijo}, since the descent direction $\nu_k(t)$ was computed using a \emph{local} second order approximation, there is no guarantee that $u_{k+1}(t)=u_k(t)+\nu_k(t)$ satisfies $g(u_{k+1})<g(u_k)$. To ensure a monotonically convergent sequence, we therefore define the next estimate
\begin{equation}
    u_{k+1}(t)=u_k(t)+\gamma_k\nu_k(t),
\end{equation}
where the step size $\gamma_k\in(0,1]$ is chosen using a a backtracking linesearch to enforce the Armijo rule
\begin{equation}\label{eq:Armijo}
    g(u_k+\gamma_k\nu_k)\leq g_k(u_k)-\alpha\gamma_k\;D g(u_k)\circ \nu_k.
\end{equation}
Details on how to compute $D g(u_k)\circ \nu_k$ are also provided in \cref{app:LQR}. A possible issue with the backtracking linesearch is that, if the Armijo rule is only satisfied for very small $\gamma$, the algorithm may perform a large number of checks before selecting a suitable step size.\medskip

To prevent unnecessary calculations, we introduce a heuristic that upper bounds the initial value of the backtracking linesearch whenever the norm of the descent direction is too large. To this end, we note that, although $x_{k+1}\in\mathcal X$ is obtained from the projection operator \eqref{eq:simp_Proj}, the LQR problem \eqref{eq:update} approximates the state update as \begin{equation}\label{eq:approx_state}
    x_{k+1}\approx x_k+\gamma_k z_k.
\end{equation}
Since \eqref{eq:opt_cstr} is a closed quantum system such that $\|x_{k+1}(t)\|=\|x_k(t)\|=\|x_0\|,~\forall t\in[0,T]$, the approximation \eqref{eq:approx_state} is valid only if $\gamma_k\|z_k(t)\|$
is ``sufficiently small'' $\forall t\in[0,T]$. This motivates the step size upper bound
\begin{equation}
    \gamma_k\leq\frac{\delta\| x_0\|}{\displaystyle\max_{t\in[0,T]}\|z_k(t)\|},
\end{equation}
where $\delta\in(0,1)$ ensures that the norm of the update is, at most, comparable to the norm of the state. A reasonable heuristic for this upper bound is $\delta=0.6$. Note that, once $\|z_k(t)\|\leq\delta\|x_0\|,$ $\forall t\in[0,T]$, i.e. once the updates are sufficiently small, we return to the full range $\gamma_k\in(0,1]$.\medskip

Although PRONTO only guarantees monotonic convergence to a local minimum, it is worth noting that there exists a large class of quantum optimal control problems for which all local minima share the same cost \cite{Rabitz2004}.

\begin{algorithm}[H]
\caption{~~Q-PRONTO}\label{alg:Q-PRONTO}
\begin{algorithmic}[1]
  \vspace{5 pt}\Statex \begin{center}
          \textbf{Initialization}\end{center}
      \State $Dg = -10\textrm{tol}$
          \Statex \emph{Initial Control Guess}
          \State $u_k(t)=u_0(t)$
          \Statex \emph{Solve \sch Equation and Compute Cost}
  \State $\xi_k\gets\mathcal P(u_k(t))$\Comment{F.i.t. \eqref{eq:simp_Proj} }
      \State $g_k\gets h(\xi_k)$
      \Statex \begin{center}
          \textbf{Main Loop}
      \end{center}
      \While{$-Dg\geq \textrm{tol}$} 
          \Statex \emph{Compute Linear-Quadratic Approximation}
      \State $[A_k(t),B_k(t),q_k(t),r_k(t),\pi_k]\gets[\eqref{eq:LinDyn}-\eqref{eq:FirstOrder}]$\vspace{5 pt}
      \Statex \emph{Perform Newton Step}
      \State Compute $\chi_k(t)$ \Comment{B.i.t \eqref{eq:adjoint}}
      \State $[Q_k(t),S_k(t),R_k(t),\Pi_k]\gets$\eqref{eq:Newton}
      \State Compute $[K_o(t),v_o(t)]$ \Comment{B.i.t \eqref{eq:DRE}}\vspace{5 pt}
      \Statex \emph{Perform quasi-Newton Step}
      \If{\eqref{eq:DRE} failed to converge}
      \State $[Q_k(t),S_k(t),R_k(t),\Pi_k]\gets$\eqref{eq:quasi_Newton}
      \State Compute $[K_o(t),v_o(t)]$ \Comment{B.i.t \eqref{eq:DRE}}
      \EndIf\vspace{5 pt}
      \Statex \emph{Compute Descent Direction and Cost Gradient}
      \State Compute $[z_k(t),\nu_k(t),\eta_k(T)]$\Comment{F.i.t. \eqref{eq:update}}
      \State $Dg\gets\pi_k^Tz_k(T)+ \eta_k(T)$\vspace{5 pt}
      \Statex \emph{Apply Armijo Rule for Step Size Selection}
      \State $\gamma_k\gets\min(\,1\,,\,\delta\| x_0\|/\max(\|z_k(t)\|)\,)$
      \State $\xi_{k+1}\gets\mathcal P(u_k(t)+\nu_k(t))$\Comment{F.i.t. \eqref{eq:simp_Proj} }
      \State $g_{k+1}\gets h(\xi_{k+1})$
      \While{$g_{k+1}>g_k-\alpha  Dg\circ\gamma_k$}
      \State $\gamma_k\gets\beta\gamma_k$
      \State $\xi_{k+1}\gets\mathcal P(u_k(t)+\gamma_k\nu_k(t))$\Comment{F.i.t. \eqref{eq:simp_Proj} }
      \State $g_{k+1}\gets h(\xi_{k+1})$
      \EndWhile\vspace{5 pt}
      \Statex \emph{Proceed to Next Iteration}
      \State $u_k(t)\gets u_k(t)+\gamma_k\nu_k(t)$
  \State $\xi_k\gets\xi_{k+1}$
      \State $g_k\gets g_{k+1}$
  \EndWhile\Statex \begin{center}
          \textbf{Output}\end{center}
      \Statex \emph{Final Control Solution}
\State \textbf{return} $u_k(t)$
\Statex \hrulefill
\Statex F.i.t. = Integrate forward in time
\Statex B.i.t. = Integrate backward in time
\end{algorithmic}
\end{algorithm}

\section{Example: qubit\\ state-to-state control}\label{sec:NumEx}
We now apply PRONTO to a canonical quantum control problem and compare it to one of the leading quantum control techniques, the Krotov method \cite{Krotov1,Krotov2}.\smallskip

\noindent Specifically, we wish to perform a state-to-state transition $\ket0\to\ket1$ on a qubit evolving under the Schr\"odinger equation
\begin{equation}\label{eq:bilinear_exp}
    \ket{\dot\psi}=-i(\hat H_0+u\hat H_1)\ket{\psi}, \quad  \ket{\psi(0)}=\ket{0},
\end{equation}
where $\hat H_0=-\frac\omega2\hat\sigma_z$, $\hat H_1=\hat\sigma_x$, $\hat\sigma_i$ are the usual Pauli matrices and $\omega=1$. To achieve this goal while limiting the fluence of the control input $u(t)$, we minimize the cost function
\begin{equation}\label{eq:cost_2spin}
    \tfrac12 \bra{\psi(T)}\hat P_{\neg 1}\ket{\psi(T)}+\int_0^T\frac {\vartheta(t)} 2 \|u(t)\|^2dt,
\end{equation}
where $\hat P_{\neg 1}=I-\oprod11= \oprod00$ is a projection operator, $\vartheta(t)>0,~\forall t\in[0,T]$ is a time-varying penalty on the energy of the control input, and $T=5~\mathrm{s}$ is the control horizon. Using the bijective mapping in \eqref{eq:bijective} and \eqref{eq:operator_bijective}, this optimization problem can then be rewritten as 
\begin{subequations}\label{eq:qubit_ocp}
\begin{eqnarray}
    \displaystyle \min~&&\tfrac12 x(T)^T\calproj{\neg 1}x(T)+\int_0^T\frac {\vartheta(t)} 2\|u(t)\|^2dt\\
    \textrm{s.t.}~&& \dot x=(H_0+u H_1)x,\quad x=x_0.
    \end{eqnarray}
\end{subequations}
To completely define the optimal control problem, we now have to specify $\vartheta(t)$. For ease of comparison\footnote{Unfortunately, the optimal control problem featured in \cite{KrotovToolkit2019} is different from \cref{eq:qubit_ocp}. The Krotov method penalizes the control update, i.e. $\int (u_k(t)-u_{k-1}(t))^2dt$, to achieve convergence, whereas the Bolza-type cost functional \eqref{eq:ocp_original} penalizes the control effort, i.e. $\int u(t)^2dt$, as in \cite{DAlessandro2001,Grivopoulos2004}. See \cite[Chap. 1]{bolzamayer_book} for further discussion on Bolza-type optimal control formulations.} with the benchmark in the Krotov Package, we assign the same input weight function featured in \cite{KrotovToolkit2019}, i.e.
\begin{equation}\label{eq:qubitcontrolcost}
    \vartheta(t)=\left\{\begin{array}{ll}
        \frac{1+\epsilon}{\mathcal B_{.6}(t)+\epsilon} & \forall t\in[0,0.3], \\
        1 & \forall t\in(0.3,4.7), \\
        \frac{1+\epsilon}{\mathcal B_{.6}(5-t)+\epsilon} & \forall t\in[4.7,5], 
    \end{array}\right.
\end{equation}
where $\epsilon=10^{-6}$ and 
\begin{equation}\label{eq:blackman}
    \mathcal B_{.6}(t)=\frac12\left(.84-\cos\left(\frac {2\pi t}{.6}\right)+.16\cos\left(\frac {4\pi t}{.6}\right)\right)
\end{equation}
is a Blackman window of length $0.6$. Note that $\epsilon>0$ ensures that $\vartheta(t)$ remains bounded for $t=0$ and $t=5$. The control cost $\vartheta(t)$ in \eqref{eq:qubitcontrolcost} is high at the beginning and end, and constant throughout the majority of the control time interval. This is used to ensure that the input starts at zero and ends at zero, \change{thereby making} the control input \change{better suited for hardware implementation}.

Finally, to initialize our iterative solver, we use the same initial guess \change{featured in} 
\cite{KrotovToolkit2019}, i.e.
\begin{equation}\label{eq:initial_guess}
    u_0(t)=\left\{\begin{array}{ll}
        0.2\, \mathcal B_{.6}(t) & \forall t\in[0,0.3], \\
        0.2 & \forall t\in(0.3,4.7), \\
        0.2\, \mathcal B_{.6}(5-t) & \forall t\in[4.7,5].
    \end{array}\right.
\end{equation}

\subsection{Benchmark Comparison, Part I}\label{ssec:Initial_converge}

\begin{figure}[t]
    \centering
    \includegraphics[width=0.5\textwidth]{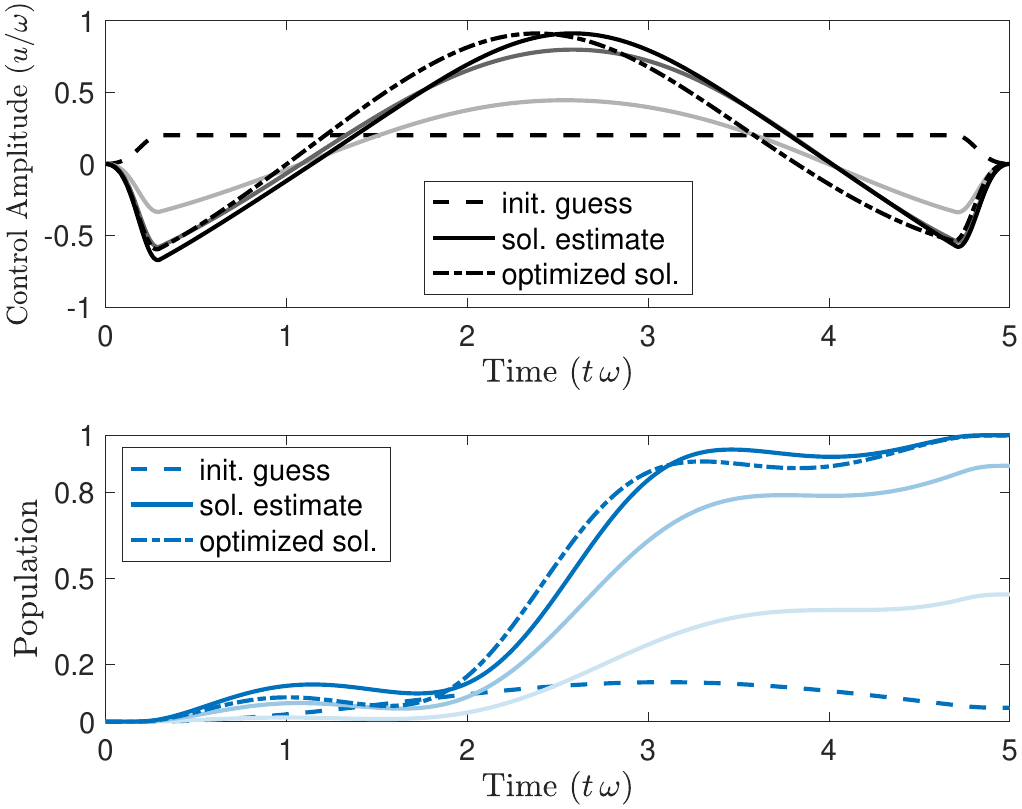}
    \caption{\change{\textbf{Top:} Control input $u(t)$ and \textbf{Bottom:} Population $P_1(t) = |\iprod{\psi(t)}{1}|^2$ for the qubit benchmark comparison. The dashed lines are the initial guess, the solid lines are the solution estimate satisfying $\texttt{tol}\leq10^{-2}$, and the dotted lines are the solution obtained using Krotov. Intermediate iterations of PRONTO are represented using semi-transparent lines.}\label{fig:result}}
\end{figure}

\begin{figure}[t]
    \centering
    \includegraphics[width=0.5\textwidth]{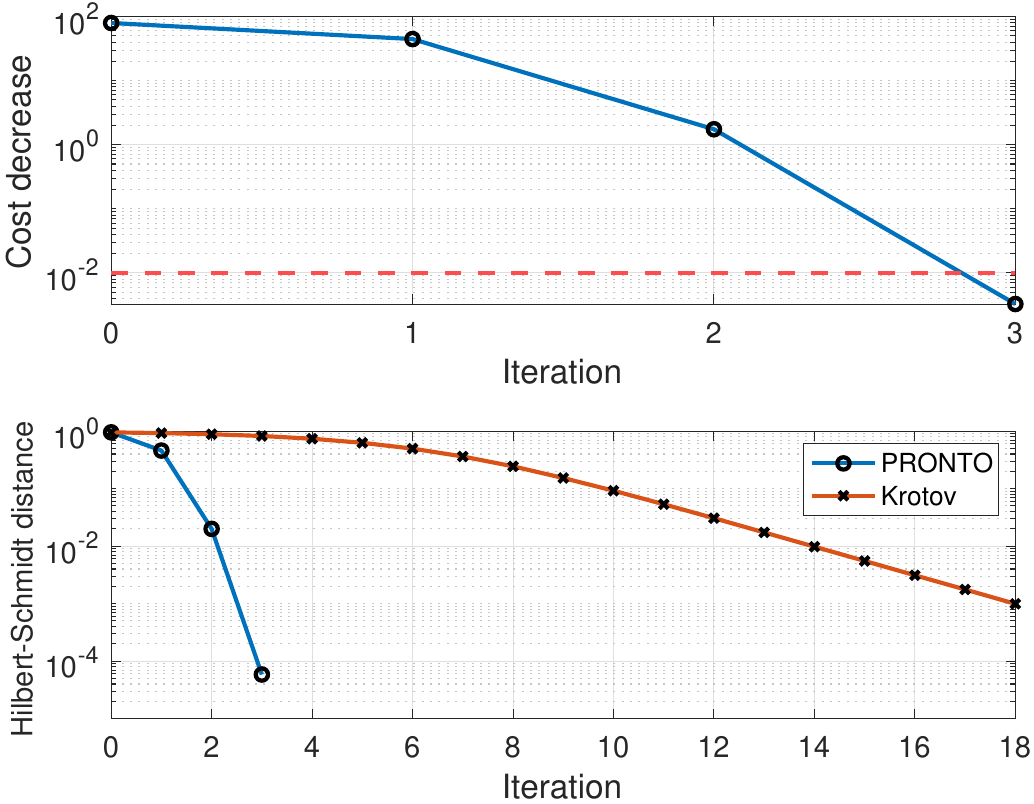}
    \caption{\textbf{Top:} Value of the cost decrease $-Dg(u_k) \circ \nu_k$ at each iteration of PRONTO. The red dashed line is the exit condition $\texttt{tol}=10^{-2}$.
    \textbf{Bottom:} Value of the squared Hilbert-Schmidt distance $1-|\iprod{\psi(T)}1|^2$ at each iteration of PRONTO and of the Krotov method. 
    \label{fig:descent} }
\end{figure}

\change{To perform an initial comparison between PRONTO and the Krotov method implemented in \cite{KrotovToolkit2019}, we begin by solving} 
the quantum optimal control problem \eqref{eq:qubit_ocp} using Algorithm \ref{alg:Q-PRONTO} \change{subject to $\texttt{tol}=10^{-2}$.}\smallskip

Figure \ref{fig:result} illustrates the behavior of the control input (top panel) and the population dynamics (bottom panel) \change{at each iteration}. In the top panel it is evident that the control amplitude \change{remains bounded and} does not contain any discontinuities
.\smallskip

Figure \ref{fig:descent} (Top) illustrates the value of the exit condition $-Dg(u_k)\circ \nu_k\leq\texttt{tol}$ at each iteration. The method meets the desired tolerance after just 3 iterations. In Figure \ref{fig:descent} (Bottom) we compare PRONTO to the Krotov method~\cite{KrotovToolkit2019} by plotting the value of the Hilbert-Schmidt distance $1-|\iprod{\psi(T)}1|^2$ obtained by each method at each iteration. For this particular example, all PRONTO iterations ended up using the Newton step as opposed to the quasi-Newton step. Although the two methods converge to similar results, the convergence rate is quadratic in the case of PRONTO and linear in the case of Krotov. This is not surprising since the former is a Newton method, whereas the latter is a gradient descent method. 

\begin{figure}
    \centering
    \includegraphics[width=0.5\textwidth]{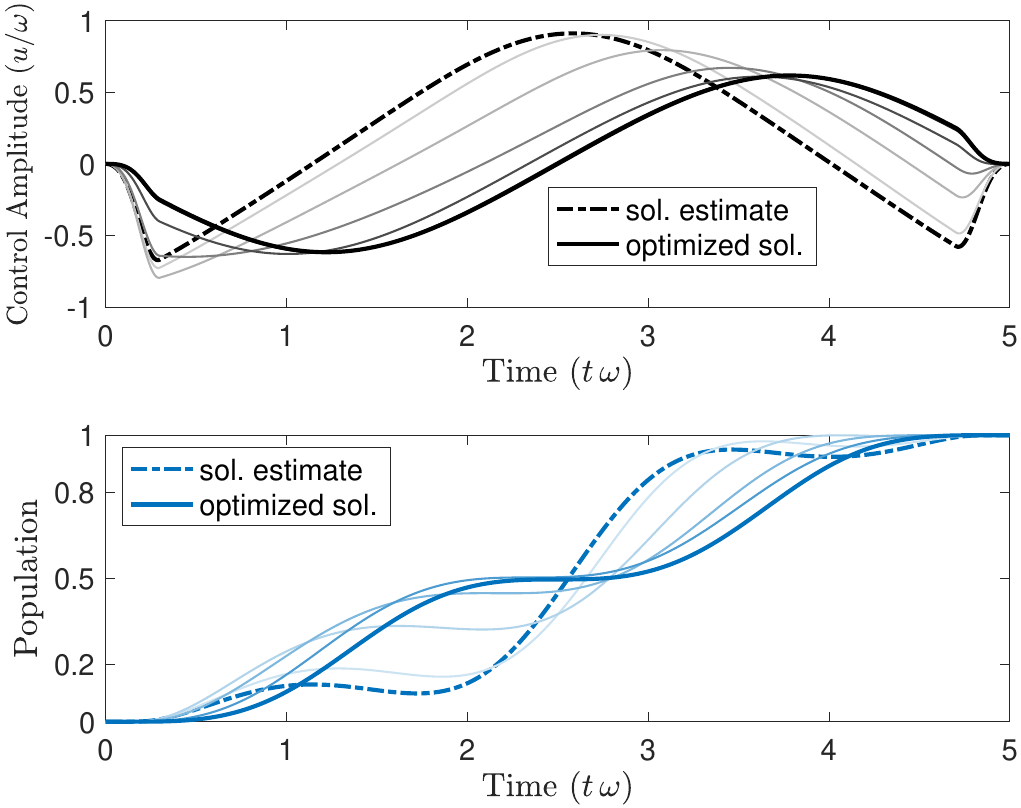}
    \caption{\change{\textbf{Top:} Control input $u(t)$ and \textbf{Bottom:} Population $P_1(t) = |\iprod{\psi(t)}{1}|^2$ for the qubit benchmark comparison. The dash-dotted lines denote the solution estimate satisfying $\texttt{tol}\leq10^{-2}$ and the solid lines denote the optimized solution satisfying $\texttt{tol}\leq10^{-8}$. Intermediate iterations of PRONTO are represented using semi-transparent lines.}}\label{fig:result_final}
\end{figure}

\begin{figure}[h]
    \centering
    \includegraphics[width=0.5\textwidth]{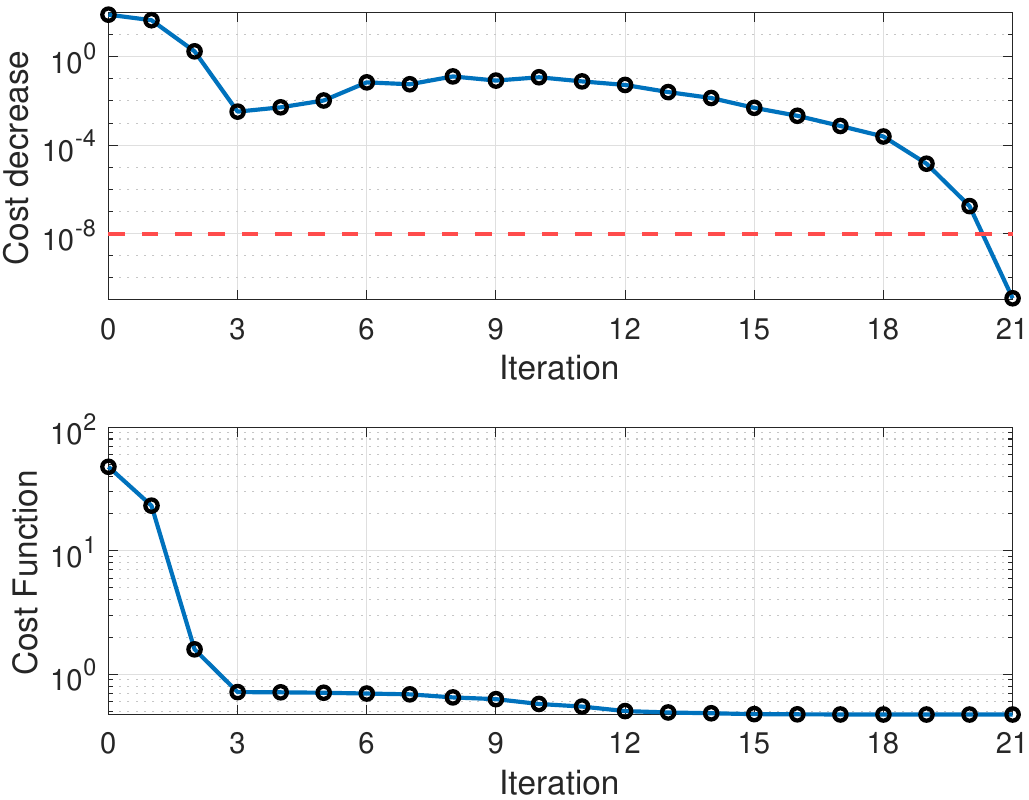}
    \caption{\change{\textbf{Top:} Value of the cost decrease $-Dg(u_k) \circ\nu_k$ at each iteration of PRONTO. The red dashed line is the exit condition $\texttt{tol}=10^{-8}$. 
    \textbf{Bottom:} Value of the cost function \eqref{eq:cost_2spin} at each iteration of PRONTO.}}\label{fig:descent_final}
\end{figure}

\change{\subsection{Benchmark Comparison, Part II}}
\change{We now continue to solve the quantum optimal control problem \eqref{eq:qubit_ocp} up to $\texttt{tol}=10^{-8}$. Intuitively, if the estimates obtained in the previous section were sufficiently close to the solution of \eqref{eq:qubit_ocp}, the method would only require a few additional iterations to meet the desired tolerance and there would be no perceivable changes to the estimate. Instead, something curious happens: the method departs from the trajectory obtained in \cite{KrotovToolkit2019} and converges to an entirely different solution.}\smallskip

\change{Figure \ref{fig:result_final} illustrates the behavior of the control input (top panel) and the population dynamics (bottom panel) at every third iteration of Algorithm \ref{alg:Q-PRONTO}. Looking at the population dynamics, we note that every iteration achieves the target objective $1-|\iprod{\psi_k(T)}1|^2\approx0$. Looking at the control input, we note that the overall amplitude of $u_k(t)$ tends to decrease at every iteration. This can be interpreted as PRONTO initially moving in the direction that minimizes the cost function (thereby obtaining a solution estimate similar to Krotov), but then refining its solution to \emph{also} minimize the control effort. This is not surprising since the quantum optimal control problem \eqref{eq:qubit_ocp} penalizes both the Hilbert-Schmidt distance \emph{and} the fluence of the control input, as opposed to the Krotov method, which only penalizes the Hilbert-Schmidt distance.}\smallskip

\change{Figure \ref{fig:descent_final} illustrates the value of the exit condition $-Dg(u_k)\circ \nu_k\leq\texttt{tol}$ (top panel) and of the cost function \eqref{eq:ocp_Qcost} (bottom panel) at each iteration. The cost function is monotonically decreasing and is lower-bounded by the value at the local minimizer. Quadratic convergence to the solution is achieved in proximity of the minimizer.}\smallskip

\change{In this case, the advantage of using PRONTO over Krotov is not in terms of requiring a lower computational effort, but in terms of obtaining an arguably ``better'' control input (i.e., the same output result is obtained with a smaller fluence $\int_0^T\vartheta(t) \|u(t)\|^2 dt$).}

\begin{figure}
    \centering
    \includegraphics[width=0.48\textwidth]{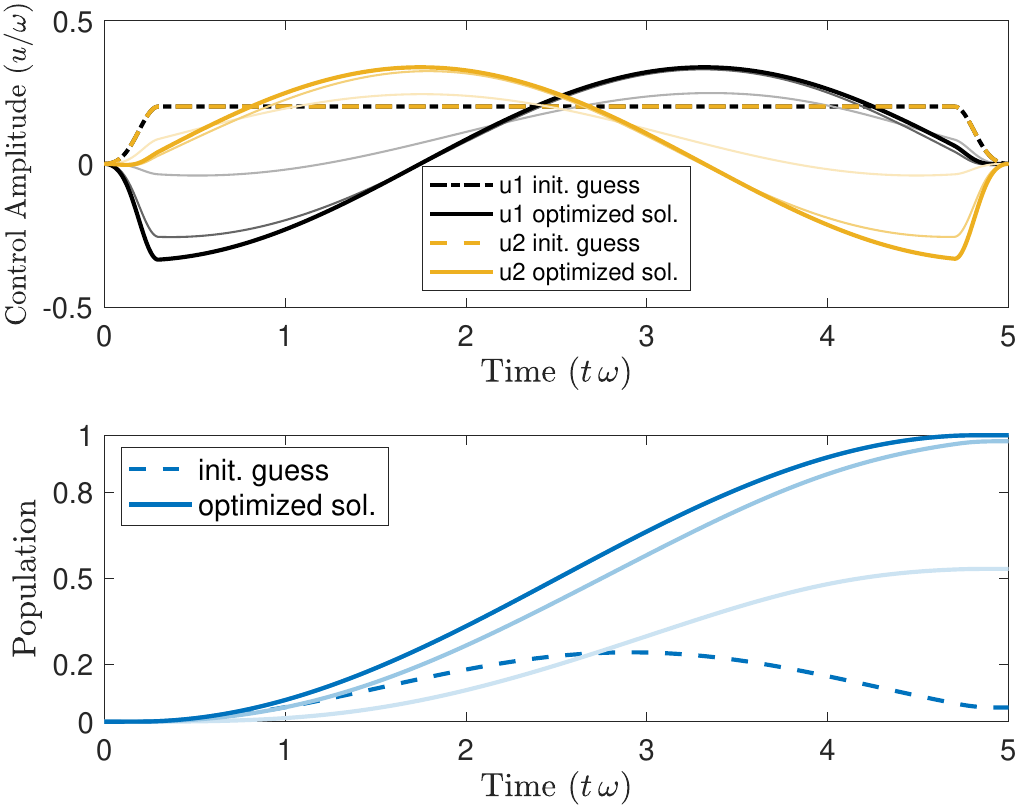}
    \caption{\change{\textbf{Top:} Control inputs $u_1(t)$, $u_2(t)$ and \textbf{Bottom:} Population $P_1(t) = |\iprod{\psi(t)}{1}|^2$ for the multi-input qubit. The dashed lines are the initial guess and the solid lines are the optimized solution satisfying $\texttt{tol}\leq10^{-8}$. Intermediate iterations are represented using semi-transparent lines.}} \label{fig:result_multi}
\end{figure}

\begin{figure}
    \centering
    \includegraphics[width=0.48\textwidth]{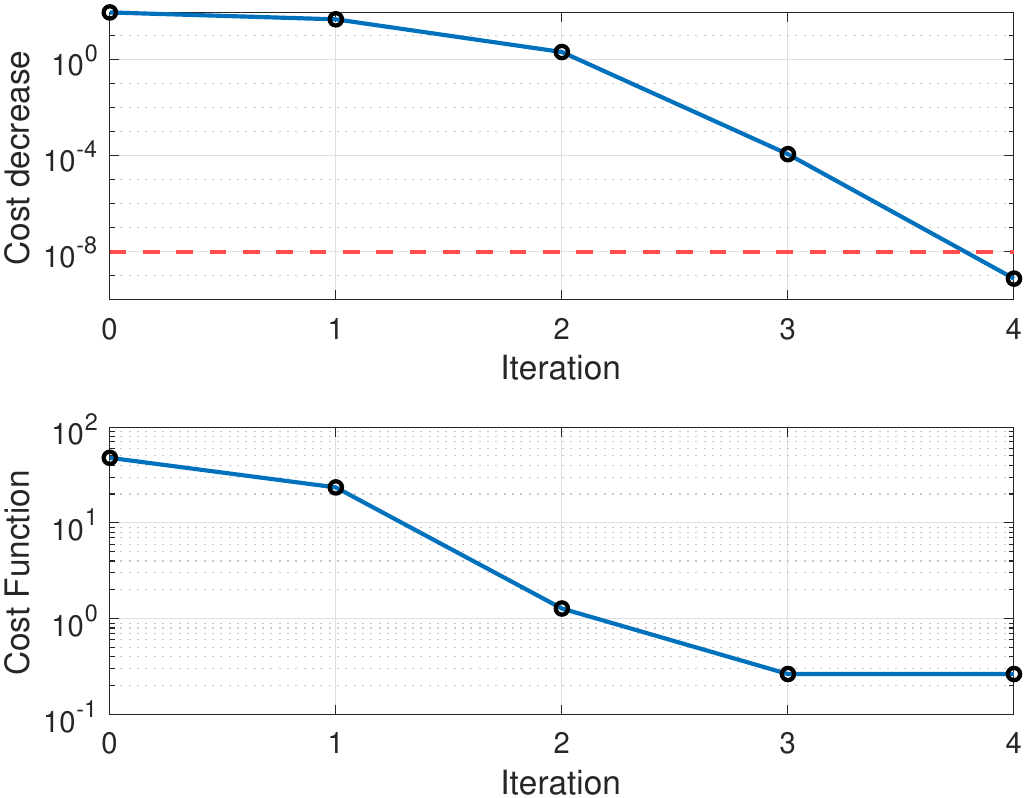}
    \caption{\change{\textbf{Top:} Value of the cost decrease $-Dg(u_k) \circ \nu_k$ at each iteration of PRONTO. The red dashed line is the exit condition $\texttt{tol}=10^{-8}$.
    \textbf{Bottom:} Value of the cost function \eqref{eq:cost_2spin} at each iteration of PRONTO.}}\label{fig:descent_multi}
\end{figure}

\change{\subsection{Multiple Control Inputs}}
\change{We now wish to perform a state-to-state transition $\ket0\to\ket1$ in the multi-input case of a qubit described by the Schr\"odinger equation
\begin{equation}\label{eq:bilinear_exp_2}
    \hbar\ket{\dot\psi}=-i(\hat H_0+u_1\hat H_1+u_2\hat H_2)\ket{\psi}, \quad  \ket{\psi(0)}=\ket{0},
\end{equation}
where $\hat H_2 =\hat\sigma_y$ and all other parameters are the same as \eqref{eq:bilinear_exp}. Given $u=[u_1, u_2]^T$, the optimization problem \eqref{eq:qubit_ocp} becomes
\begin{subequations}\label{eq:qubit_ocp_2}
\begin{eqnarray}
    \displaystyle \min~&&\tfrac12 x(T)^T\calproj{\neg 1}x(T)+\int_0^T\frac {\vartheta(t)} 2\|u(t)\|^2dt\\
    \textrm{s.t.}~&& \dot x=(H_0+u_1 H_1+u_2 H_2)x,\quad x=x_0.
    \end{eqnarray}
\end{subequations}
The initial guess \eqref{eq:initial_guess} is used for both $u_1$ and $u_2$. The quantum optimal control problem \eqref{eq:qubit_ocp_2} is solved using Algorithm \ref{alg:Q-PRONTO} subject to $\texttt{tol}=10^{-8}$.}\smallskip

\change{Figure \ref{fig:result_multi} illustrates the behavior of the control input (top panel) and the population dynamics (bottom panel) at each iteration.  In this example, it is interesting to note that the shape of $u_1(t)$ is similar to the one found in the previous section, but the amplitude is lower due to the contribution of $u_2(t)$.} \smallskip

\change{Figure \ref{fig:descent_multi} illustrates the value of the exit condition $-Dg(u_k)\circ \nu_k\leq\texttt{tol}$ at each iteration. The method quadratically converges to the optimizer in 4 iterations. Once again, the cost function decreases monotonically and quadratically converges to a lower bound.}\bigskip

\section{Conclusions}
This paper specialized the Projection Operator-based Newton method for Trajectory Optimization (PRONTO) to quantum control problems. The method is guaranteed to be monotonically convergent at all times and features a quadratic convergence rate in proximity of local minima.

There are many directions for future work.
The original PRONTO method~\cite{PRONTO2002,qPRONTO} included a ``regulator'' that is not featured in the quantum extension presented here. Thus, an important technical extension of the quantum PRONTO is to design a quantum-specific regulator to be used in the projection operator. This will likely enable faster convergence for higher-dimensional quantum systems. Additional extensions include the design of unitary gates, control of open quantum systems, and in-depth comparisons with existing optimization methods. Eventually, we plan to release an open source toolkit/package for quantum control using PRONTO.\smallskip

\noindent {\em Acknowledgements}
The authors would like to thank Liang-Ying Chih and Murray Holland for the helpful discussions. The work by Jieqiu Shao and Marco M. Nicotra is supported by the NSF QII–TAQS award number 1936303.


\bibliography{project_bib.bib}
\vspace{2em}

\appendix

\section{Fr\'echet Derivatives}\label{app:derivatives}
This appendix shows how to obtain the LQR problem \eqref{eq:LQR} by formally computing the first and second Fr\'echet derivatives featured in \eqref{eq:simp_LQR}.\bigskip

\noindent \textbf{First Derivatives}\\
Using the chain rule, we obtain
\begin{equation}\label{eq:first_der}
     D g(u_k)\circ \nu=Dh(\mathcal{P}(u_k))\circ D\mathcal{P}(u_k)\circ \nu.
\end{equation}
Following from the definition of the projection operator \eqref{eq:simp_Proj}, we have $\mathcal{P}(u_k)=\xi_k$. Moreover, its local derivative $\zeta(t)=(z(t),v(t))$ satisfies
\[
        D\mathcal{P}(u_k)\circ \nu=\zeta: \left\{\begin{array}{l}
     \dot{z}=A_k(t) z +B_k(t)v, ~~z(0)=0\\
         v = \nu,
    \end{array}\right.
\]
with $A_k(t)$ and $B_k(t)$ defined in \eqref{eq:LinDyn}. Thus, the first derivative can be rewritten as
\begin{equation}
     D g(u_k)\circ \nu=Dh(\xi_k)\circ \zeta.
\end{equation}
It then follows from  \eqref{eq:cost} that
\[
    Dh(\xi_k)\circ\zeta=\pi_k^T z(T)+\int_0^Tq_k^T(\tau)z(\tau)+r_k^T(\tau)\nu(\tau)d\tau,
\]
with \begin{equation}
\begin{array}{rl}
    q_k(t)&=\nabla_xl(x_k(t),u_k(t)),\\
    r_k(t)&=\nabla_ul(x_k(t),u_k(t)),\\
    \pi_k&=\nabla_x m(x_k(T)),
\end{array}
\end{equation}
which translate into \eqref{eq:FirstOrder} once all the partial derivatives are evaluated.\bigskip

\noindent\textbf{Second Derivatives}\\
By applying the chain rule to \eqref{eq:first_der}, we obtain
\[
\begin{array}{rl}
     D^2 g(u_k)\!\circ\! (\nu,\nu)=&\!D^2h(\mathcal{P}(u_k))\!\circ\! (D\mathcal{P}(u_k)\!\circ\! \nu,D\mathcal{P}(u_k)\!\circ\! \nu)\\&
     \!\!\!\!+\,Dh(\mathcal{P}(u_k))\!\circ\! D^2\mathcal{P}(u_k)\!\circ\!(\nu,\nu).
\end{array}
\]
Substituting $\mathcal{P}(u_k)=\xi_k$ and $D\mathcal{P}(u_k)\circ \nu=\zeta$ then implies
\begin{equation}
\begin{array}{rl}
 D^2 g(u_k)\circ (\nu,\nu)=& D^2h(\xi_k)\circ (\zeta,\zeta)+\ldots\\&Dh(\xi_k)\circ D^2\mathcal{P}(u_k)\circ(\nu,\nu).
\end{array}
\end{equation}
It then follows from  \eqref{eq:cost} that the first term satisfies
\[
\begin{array}{rl}
     D^2h(\xi_k)\circ(\zeta,\zeta)=\! & z(T)^T\Pi_k z(T)+\ldots \\
     &\displaystyle \int_0^T\!\begin{bmatrix}z(\tau)\\\nu(\tau)\end{bmatrix}^{\!T}\!\!\begin{bmatrix}Q_k(\tau) & S_k(\tau)\\S_k^T(\tau)&R_k(\tau)\end{bmatrix}\!\!\begin{bmatrix}z(\tau)\\\nu(\tau)\end{bmatrix}
    d\tau,
\end{array}
\]
with 
\begin{equation}
\begin{array}{rl}
    Q_k(t)&=\nabla^2_{xx}l(x_k(t),u_k(t))\\
    S_k(t)&=\nabla^2_{xu}l(x_k(t),u_k(t))\\
    R_k(t)&=\nabla^2_{uu}l(x_k(t),u_k(t))\\
    \Pi_k&=\nabla^2_{xx} m(x_k(T)),
\end{array}
\end{equation}
which yield \eqref{eq:quasi_Newton} when evaluated.\medskip

As for the second term, we note that the second derivative of the projection operator \eqref{eq:simp_Proj} yields
\[
        D^2\mathcal{P}(u_k)\circ (\nu,\nu)= \left\{\begin{array}{l}
    \dot{y}=A_k(t) y +B_k(t)w+\phi_k(t)\\
         w = 0,
    \end{array}\right.
\]
with $y(0)=0$ and
\[
    \phi_k(t)=
    \sum_{i=1}^m\nu_i(t)\mathcal H_i(t)z(t)+\displaystyle\sum_{i,j=1}^m\nu_i(t)\nu_j(t)\mathcal H_{ij}(t)x_k(t).
\]
Given the state transition matrix
$\Phi(t,\tau)$ satisfying
\begin{equation}
    \frac\partial{\partial t}\Phi(t,\tau)=A(t)\Phi(t,\tau),
\end{equation}
we can write
\begin{equation}
    y(t)=\int_0^t\Phi(t,s)\phi_k(s)ds.
\end{equation}
Thus, we have
\[
\begin{array}{l}
    Dh(\xi_k)\circ D^2\mathcal{P}(u_k)\circ(\nu,\nu)\\ \displaystyle=\pi_k^T y(T)+\int_0^Tq_k^T(\tau)y(\tau)d\tau\\\displaystyle=
    \pi_k^T y(T)+\int_0^Tq_k^T(\tau)\int_0^\tau\Phi(\tau,s)\phi_k(s)ds\,d\tau\\
    \displaystyle=\pi_k^T y(T)+\int_0^T\int_s^T q_k^T(\tau)\Phi(\tau,s)d\tau\:\phi_k(s)ds\\
    \displaystyle=\int_0^T\left(\pi_k^T \Phi(T,s)+\int_s^T q_k^T(\tau)\Phi(\tau,s)d\tau\right)\phi_k(s)ds
\end{array}
\]
Given
\begin{equation}
    \chi_k(s)=\Phi(T,s)^T \pi_k+\int_s^T \Phi(\tau,s)^T q_k(\tau)d\tau,
\end{equation}
it follows from the properties of state transition matrices that $\chi:[0,T]\to\mathbb R^{2n}$ can be obtained by solving the differential equation \eqref{eq:adjoint}. Thus, we obtain
\[
    Dh(\xi_k)\circ D^2\mathcal{P}(u_k)\circ(\nu,\nu) = \int_0^T\chi_k^T(s)\phi_k(s)ds,
\]
where it is possible to show that
\[
    \chi_k^T(s)\phi_k(s)=\!\begin{bmatrix}z(s)\\\nu(s)\end{bmatrix}^{\!T}\!\!\begin{bmatrix}0 & \tilde S_k(s)\\\tilde S_k^T(s)&\tilde R_k(s)\end{bmatrix}\!\!\begin{bmatrix}z(s)\\\nu(s)\end{bmatrix},
\]
with $\tilde R_k(\cdot)$ and $\tilde S_k(\cdot)$ the same as in \eqref{eq:Newton}.

\section{Linear-Quadratic Optimal Control}\label{app:LQR}
The optimal control problem \eqref{eq:LQR} is a special class of trajectory optimization problems for which it is possible to compute an explicit solution \cite{Opt_ctrl}. To do so, we first solve the backwards in time Differential Riccati Equation
\begin{equation}\label{eq:DRE}
    \left\{\begin{array}{rll}
    -\dot P =&A^T_kP\!+\!PA_k-K_o^TR_kK_o+Q_k,~~ & P(T)\!=\!\Pi_k,\\
    -\dot p=&(A_k-B_kK_o)^Tp-K_o^Tr_k+q_k,& p(T)=\pi_k,\\
    K_o =& R^{-1}_k(B_k^TP+S_k^T),\\
    v_o = & R_k^{-1}(B^T_kp+r_k).
    \end{array}\right.
\end{equation}
Having solved for $K_o(t)$ and $v_o(t)$, it is then possible to obtain $\nu_k(t)$ by solving
\begin{equation}\label{eq:update}
    \left\{\begin{array}{rll}
    \dot \eta_k =&q_k^Tz_k+r_k^T\nu_k,~~ & \eta(0)=0,\\
    \dot z_k =&A_kz_k+B_k\nu_k,~~ & z(0)=0,\\
    \nu_k=&-v_o-K_oz_k,
    \end{array}\right.
\end{equation}
where $z_k\in\mathcal X$ is a local approximation of the state update and the running cost $\eta_k:[0,T]\to\mathbb R_{\geq0}$ is used to compute
\begin{equation}
    Dg(u_k)\circ \nu_k=\pi_k^T z_k(T)+\eta_k(T),
\end{equation} 
which is needed to a) determine the step size $\gamma_k$ via the Armijo rule \eqref{eq:Armijo}, and b) verify the exit condition for the iterative solver $-Dg(u_k)\cdot\nu_k\leq\texttt{tol}$.

\end{document}